\documentclass[12pt,a4paper]{article}
\usepackage{times}
\usepackage{amsmath}
\usepackage[dvips]{graphicx}
\begin{document}

\begin{titlepage}
\title{Novel features of diffraction at the LHC}
\author{V.~A.~Petrov, A.~V.~Prokudin, S.~M.~Troshin,
N.~E.~Tyurin\\
\small  \it Institute for High Energy Physics,\\
\small  \it Protvino, Moscow Region, 142280, Russia} \normalsize
\date{}
\maketitle

\begin{abstract}
Interest and problems in the studies of diffraction at LHC are
highlighted.
 Predictions for the global characteristics of
proton-proton interactions at the LHC energy are given.
 Potential discoveries of the antishadow scattering mode
  which is allowed in principle by unitarity and diffractive
scattering
conjugated with high--$E_T$ jets are discussed.
\end{abstract}

\end{titlepage}

\setcounter{page}{2}
\section*{Introduction}

 During  recent years CERN, DESY and FNAL have been
producing  interesting results on diffractive production in hadron
and deep-inelastic processes \cite{rev}. Discovery of hard
diffraction at CERN S$\bar p p$S \cite{cern} and  diffractive
events in the deep-inelastic scattering at HERA \cite{h1,zeus}
were among the most surprising results obtained recently.
Significant fraction of high-$t$ events among the diffractive
events  in deep-inelastic scattering and in hadron-hadron
interactions were also observed at HERA \cite{herdif} and Tevatron
\cite{tevdif} respectively. These experimental findings  have
renewed interest in the experimental and theoretical
  studies of the diffractive
production processes.

There are  many unsolved problems in soft and hard hadronic
physics which should be studied at the highest possible energies
at the LHC  and their importance should not be overshadowed by the
expectations for the new particles  in this newly opening energy
range. We consider several such problems in some details in this
note.

First of all one deals with genuinely \bf{strong interactions}\rm,
which are not corrections to the free or lowest--order dynamics
(this is the case of purely hard processes where perturbative QCD
is able (with some serious reservations, though) to make
predictions and decriptions). In this regime it is possible that
the interaction will enter the new scattering mode -- antishadow
scattering which is in principle allowed by unitarity and may be
realized in the region of the strong coupling \cite{phl}.
However, it is not necessarily realized in nature and only the
experimental studies can provide the crucial answer.

It is useful to estimate spatial extension of the diffractive
processes. From the Heisenberg uncertainty relations one gets, e.
g. for elastic scattering,
\begin{eqnarray}
\Delta x_i \Delta p_i & \geq & 1,\quad i=\parallel,\perp\nonumber\\
(\Delta p_\parallel)^2 & = & (\langle t^2 \rangle -\langle t \rangle^2)/4p^2,
\nonumber\\
4p^2 & = &s-4m^2\nonumber\\
(\Delta p_\perp)^2 & = & -\langle t \rangle +\langle t \rangle^2/4p^2,
\end{eqnarray}
and at high energies
\begin{eqnarray}
\Delta x_\|^* & \geq & \sqrt{s}/\sqrt{\langle t^2\rangle - \langle
t\rangle ^2}\nonumber\\
\Delta x_\perp & \geq & 1/\sqrt{\langle -t \rangle},
\end{eqnarray}
where $\Delta x_\|^*$ and $\Delta x_\perp$ are longitudinal and
transverse coordinate uncertainties, correspondingly in the c. m.
s., $\sqrt{s}$ is the total c. m. s. energy.
It should be noted that our formulas refer to final state momenta
 which are stochastic due to
fluctuations (quantum-mechanical) in the scattering angle
and   our definition of $(\Delta p)^2$
looks like the following:
$\langle p_\parallel\rangle = p\langle \cos(\theta)\rangle$,
$\langle p^2_\parallel\rangle = p^2\langle \cos^2(\theta)\rangle$
and then we take
as usual
\[
( \Delta p_\parallel )^2 =
 p^2(\langle\cos^2\theta\rangle - \langle \cos\theta\rangle^2);
\]
similarly for $\Delta p_\perp$, but there we know due to azimuthal
symmetry that
$\langle \vec{p}_\perp\rangle = 0$.

In diffractive processes average momentum transfers $\langle
-t\rangle$, $\langle t^2 \rangle$ depend only weakly on $s$ so we
will deal with \bf{large distances} \rm  at LHC.
For instance
\[
\Delta x_\|^*  > 40000\; \mbox{fm !}
\]

At such long distances description of the high--energy collisions
in terms of individual partons --- quarks and gluons ceases to be
adequate. We enter
 a new territory where confinement dynamics is overwhelming and some (gluon)
 field configurations become relevant degrees of freedom. In other words
 diffractive high--energy scattering deals with undulatory aspects of the
 QCD dynamics.

 This field is one of the greatest challenges to both theoretical and experimental
  high--energy physics communities.

\section{Antishadow Scattering at LHC}

 Unitarity of the scattering matrix $SS^+=1$ implies the
existence at high energies $s>s_0$ of the new scattering mode --
antishadow one. It has been described in some detail (cf.
\cite{ech} and references therein) and the most important feature
of this mode is the self-damping of the  contribution from the
inelastic channels. We argue here that
  the antishadow scattering mode could be definitely
revealed at  the LHC energy and
 provide numerical estimations based on the $U$-matrix
unitarization method
 \cite{ltkhs}.
\begin{figure}[t]
 \begin{center}
\resizebox{10cm}{!}{\includegraphics{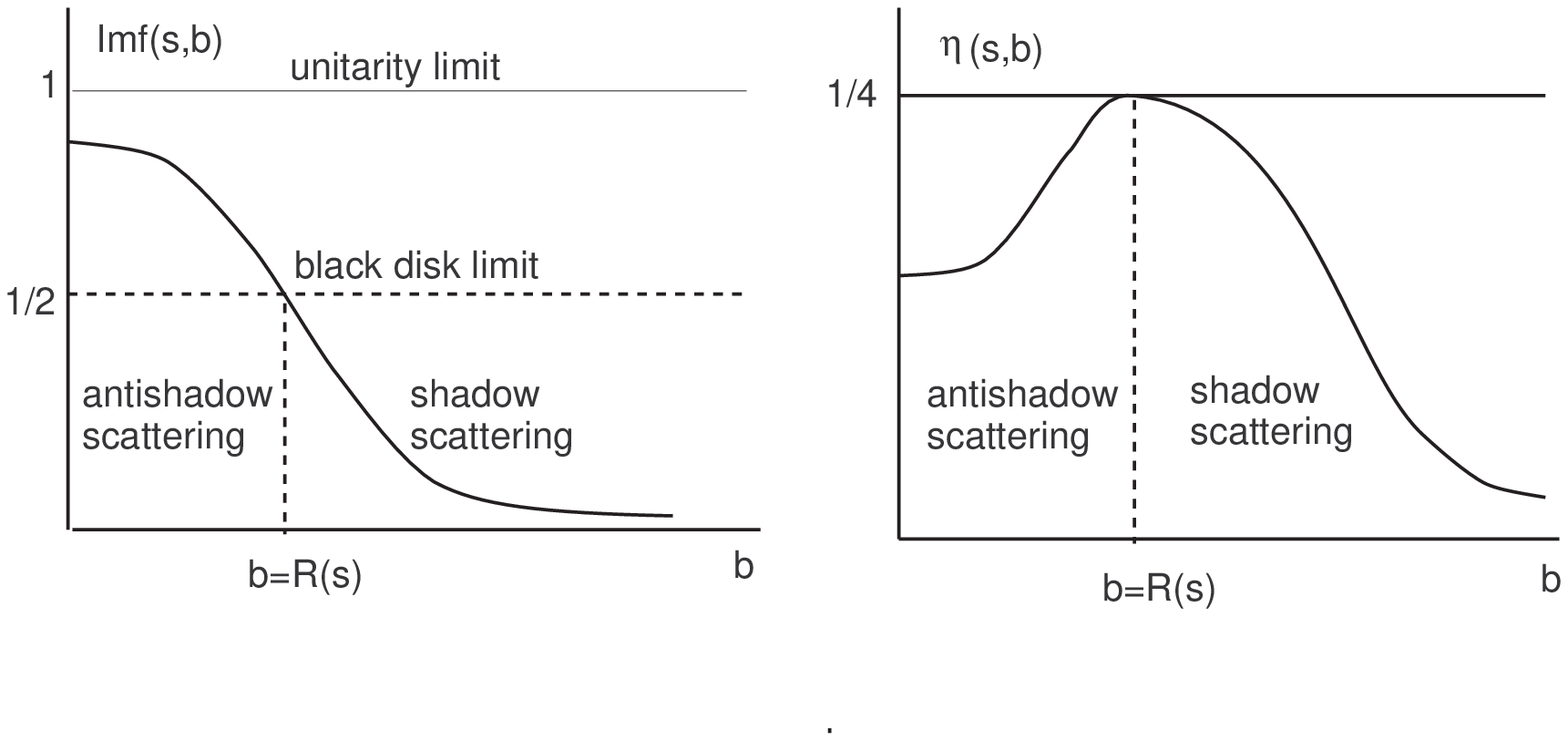}}
\end{center}
 \vspace{-1.5cm}
 \caption{Shadow and antishadow scattering regions}
 \end{figure}
In the impact parameter representation the unitarity relation
written for the elastic scattering amplitude $f(s,b)$ at high
energies has the form
\begin{equation}
Im f(s,b)=|f(s,b)|^2+\eta(s,b) \label{unt}
\end{equation}
where the inelastic overlap function $\eta(s,b)$ is the sum of all
inelastic channel contributions.
 Unitarity equation  has
two solutions for the case of pure imaginary amplitude:
\begin{equation}
f(s,b)=\frac{i}{2}[1\pm \sqrt{1-4\eta(s,b)}].\label{usol}
\end{equation}
Eikonal unitarization
\begin{equation}\label{eik}
  f(s,b)=\frac{e^{2i\delta(s,b)}-1}{2i}
\end{equation}
with pure imaginary eikonal  ($\delta=i\Omega/2$) corresponds to
the choice of the one particular solution of the unitarity
equation with sign minus.

In the $U$--matrix approach the form of the elastic scattering
amplitude in the impact parameter representation is the following:
\begin{equation}
f(s,b)=\frac{U(s,b)}{1-iU(s,b)}. \label{um}
\end{equation}
 $U(s,b)$ is the generalized reaction matrix, which is considered as an
input dynamical quantity similar to eikonal function.

Inelastic overlap function is connected with $U(s,b)$ by the
relation
\begin{equation}
\eta(s,b)=\frac{Im U(s,b)}{|1-iU(s,b)|^{2}}\label{uf}.
\end{equation}

\begin{figure}[t]
 \begin{center}
\resizebox{10cm}{!}{\includegraphics{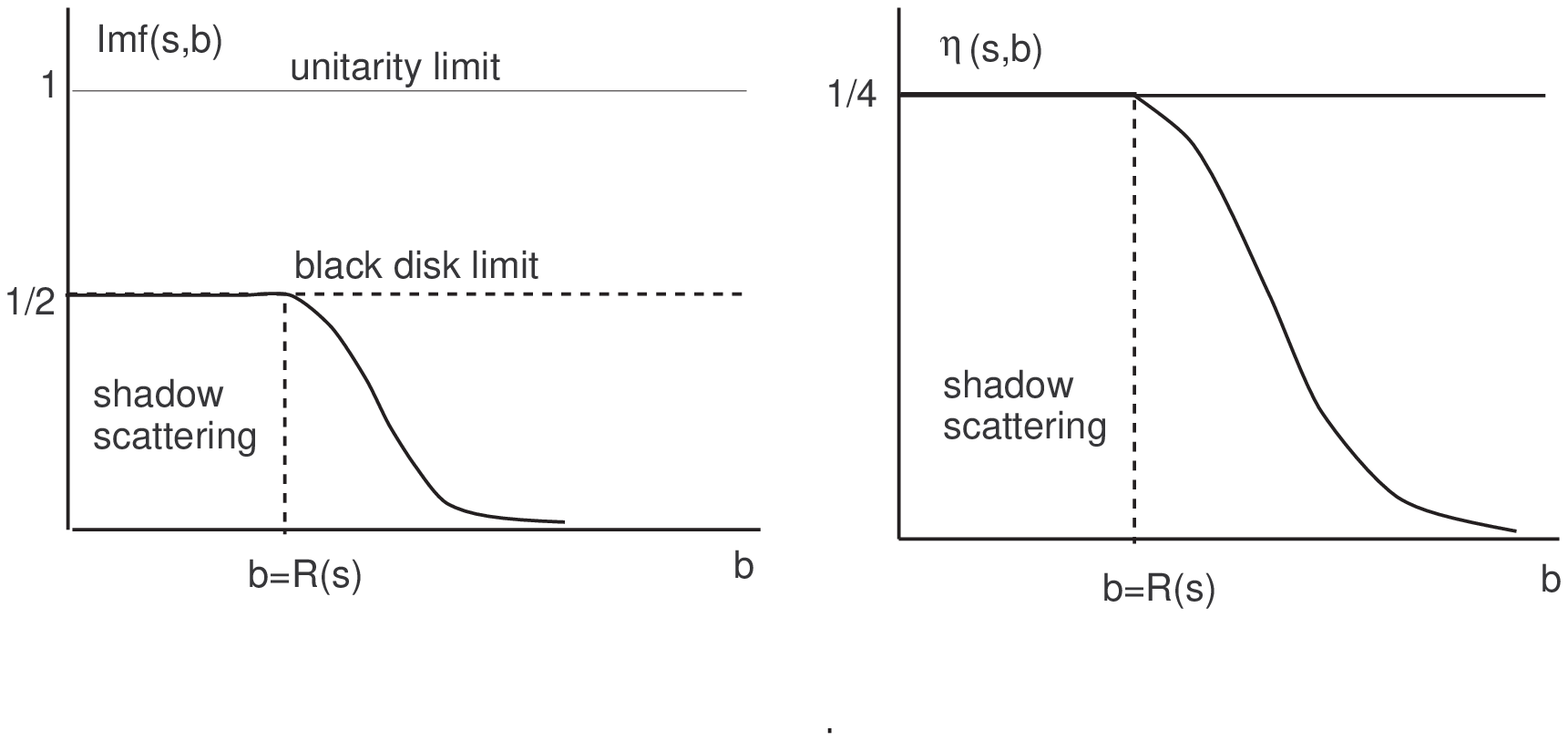}}
\end{center}
 \vspace{-1.5cm}
 \caption{Shadow scattering mode}
 \end{figure}

It is worth noting that the shadow scattering mode is  considered
usually as the only possible one. But the two solutions of the
unitarity
 equation have an equal meaning and the antishadow scattering mode
 should not be excluded.

 Appearance of the antishadow scattering mode
is completely consistent with the basic idea that the particle
production is the driving force for elastic scattering. Let us
consider the transition to the antishadow scattering mode
 \cite{phl}. With
conventional parameterizations of the $U$--matrix
 the inelastic overlap function increases with energies
at modest values of $s$. It reaches its maximum value
$\eta(s,b=0)=1/4$ at some energy $s=s_0$ and beyond this energy
the  antishadow scattering mode appears at small values of $b$.
The region of energies and impact parameters corresponding to the
antishadow scattering mode is determined by the conditions $Im
f(s,b)> 1/2$ and $\eta(s,b)< 1/4$. The quantitative analysis of
the experimental data
 \cite{pras} gives the threshold value of energy: $\sqrt{s_0}\simeq 2$ TeV.
This value is confirmed by the recent model considerations
\cite{laslo}.

Thus, the function $\eta(s,b)$ becomes peripheral when energy is
increasing. At such energies the inelastic overlap function
reaches its maximum
 value at $b=R(s)$ where $R(s)$ is the interaction radius.
So, beyond the transition threshold there are two regions in
impact
 parameter space: the central region
of antishadow scattering at $b< R(s)$ and the peripheral region of
shadow scattering at $b> R(s)$. The impact parameter dependence of
the amplitude $f(s,b)$ and inelastic channel contribution
$\eta(s,b)$ at $s>s_0$ are represented on Fig. 1.
\begin{figure}[t]
\begin{center}
  \resizebox{6cm}{!}{\includegraphics{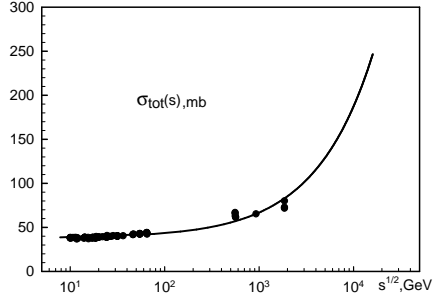}}
\end{center}
\caption{Total cross-section of $pp$--interactions, experimental data from
\cite{ezh}}
\end{figure}

The region of LHC energies is the one where antishadow scattering
 mode is to be presented.  This mode can be revealed directly measuring
 $\sigma_{el}(s)$ and $\sigma_{tot}(s)$ and not only through the
 analysis in impact parameter representation.

 Note that the impact parameter behavior of the amplitude
 and the inelastic overlap function have the form depicted on
 the Fig. 2 in case when the only shadow scattering is realized
 at the LHC energies.

For the LHC energy $\sqrt{s}= 14$ $TeV$ the model based on the
$U$--matrix form of unitariazation provides (Fig. 3)
\begin{equation}\label{s}
 \sigma_{tot}\simeq 230\; \mbox{mb}
\end{equation}
and
\begin{equation}\label{r}
\sigma_{el}/\sigma_{tot}\simeq 0.67.
\end{equation}
 Thus, the antishadow scattering mode could be discovered
at LHC by measuring $\sigma_{el}/\sigma_{tot}$ ratio which is
greater than the black disc value $1/2$ (Fig. 4).

However, the LHC energy is not in the asymptotic region yet, the
asymptotical behavior
\begin{equation}\label{tota}
  \sigma_{tot,el}\propto \ln^2 s,\;\;
 \sigma_{inel}\propto \ln s
\end{equation}
is expected at $\sqrt{s}> 100$ $TeV$.

The above predicted values for the global characteristics of $pp$
-- interactions at LHC differ from the most common predictions of
the other models. First of all total cross--section is predicted
to be twice as much of  the common predictions in the range 95-120
mb \cite{vels}
 and it even overshoots the existing cosmic ray data. However,
 extracting proton--proton cross sections from cosmic ray
 experiments is model dependent and  far from straightforward
 (see, e.g. \cite{bl} and references therein). It should be noted
 here that the large value of the total cross--section is due to
 the elastic scattering while the value of inelastic cross--section
 is about 80 mb and close to the common predictions. Therefore, the
 large value of the total cross--section does not imply the large
 background.
\begin{figure}[t]
\begin{center}
\resizebox{6cm}{!}{\includegraphics{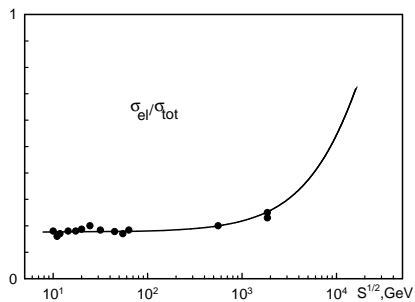}}
\end{center}
 \caption{Ratio
of elastic to total cross-section of $pp$--interactions,
experimental data from \cite{ezh}}
\end{figure}

\section{Inelastic Diffraction at LHC}
 Similarity between elastic and inelastic diffraction
in the $t$-channel approach suggests that the latter one would
have similar to elastic scattering behavior of the differential
cross-section. However, it cannot be taken for granted and   e.g.
transverse momentum distribution of diffractive events in the
deep-inelastic scattering at HERA  shows a power-like behavior
without apparent dips \cite{will}. Similar behavior was observed
also in the hadronic diffraction dissociation process at CERN
\cite{cern} where also no dip and bump structure was observed.
Angular dependence of diffraction dissociation together with the
measurements of the differential cross--section in elastic
scattering would allow to determine the geometrical properties of
elastic and inelastic diffraction, their similar and distinctive
features and  origin.

It is interesting to note that at large values of the missing mass
$M^2$ the normalized differential cross-section
$\frac{1}{\sigma_0}\frac{d\sigma_D }{ dtdM^2}$ ($\sigma_0$ is the
value of cross-section at $t=0$) will exhibit scaling behavior
\cite{angd}
\begin{equation}\label{scal}
\frac{1}{\sigma_0}\frac{d\sigma_D }{dtdM^2}=f(-t/M^2),
\end{equation}
and explicit form of the function $f(-t/M^2)$ is the following
\begin{equation}\label{ftau}
 f(-t/M^2)=(1-4\xi ^2t/M^2)^{-3}.
\end{equation}
This dependence is depicted on Fig. 5.

 The above scaling has been obtained in the model approach,
however it might have a more general meaning. Conventional
diffractive inelastic scattering predictions on the basis
of the triple-reggeon phenomenology do not exhibit $t/M^2$--scaling.

 The angular structure of
diffraction dissociation processes given by Eq. (\ref{scal}) takes
place  at high energies where  while at moderate and low energies
dip--bump structure can be presented \cite{angd}. Thus at low
energies the situation is similar to the elastic scattering, i.e.
diffraction cone and possible dip-bump structure should be present
in the region of small values
 of $t$ and behavior of the differential
cross-section will be rather complicated and incorporates
diffraction cone, Orear type (exponential behavior with
$\sqrt{-t}$) and power-like dependencies.

 At
the LHC energy the diffractive events with the masses as large as
3 TeV could be studied. It would be interesting to check this
prediction at the LHC  where the scaling and simple power-like
behavior of diffraction dissociation differential cross-section
should be observed. Observation of such behavior would confirm the
diffraction mechanism based on excitation of the complex
hadronlike  object - constituent quark. This mechanism can in
principle explain angular structure of diffraction
 in the deep - inelastic
scattering at HERA where smooth angular dependence on the thrust
transverse momentum was observed \cite{will}. If it is the case,
then diffraction in DIS at lower energies should manifest typical
soft diffractive behavior with exponential peak at small $t$ as it
does in hadronic reactions.
\begin{figure}[t]
 \begin{center}
\resizebox{8cm}{!}{\includegraphics{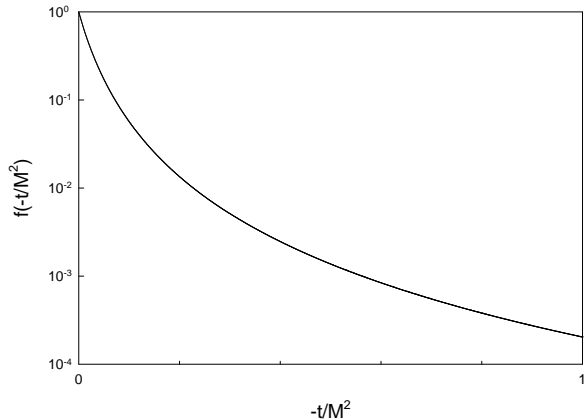}}
\end{center}
\caption{Scaling behavior of the normalized differential
cross-section $\frac{1}{\sigma_0}\frac{d\sigma }{ dtdM^2}$.}
 \end{figure}

\section{Hard and Soft Diffraction Interplay at LHC}
In principle measurements of the global characteristics, like
$\sigma_{tot}$, $\sigma_{el}$, $\sigma_{D(D)}$, $d\sigma/dt$ etc.
may be considered as a source of information on the size and shape
of the interaction region. To some extent this can be assimilated
to the famous ``inverse scattering problem'' in potential
scattering, where the problem is, roughly, to extract an unknown
potential from the ``data''
 (phase shifts).

 This stage of study is, in principle, model independent. Only after getting
 an information on the interaction region can one ask if, say, QCD is
 able to describe and explain it.

 When generic diffractive processes proceed it may happen that due to
 vacuum fluctuations some short--time perturbation will take place,
 resulting in appearing of hard scattered partons which we percept
 as hadronic jets. Such a perturbation may quite strongly influence
 the interaction region which can result in a spectacular change of
 the normal diffractive pattern.

 As an example one can consider the process (Fig. 6)
 \[
 p+p\to p+\mbox{jet}+\mbox{jet}+p,
 \]
 where two jets are safely separated from ``diffractive'' protons
 by rapidity gaps.
\begin{figure}[hbt]
 \begin{center}
\resizebox{5cm}{!}{\includegraphics{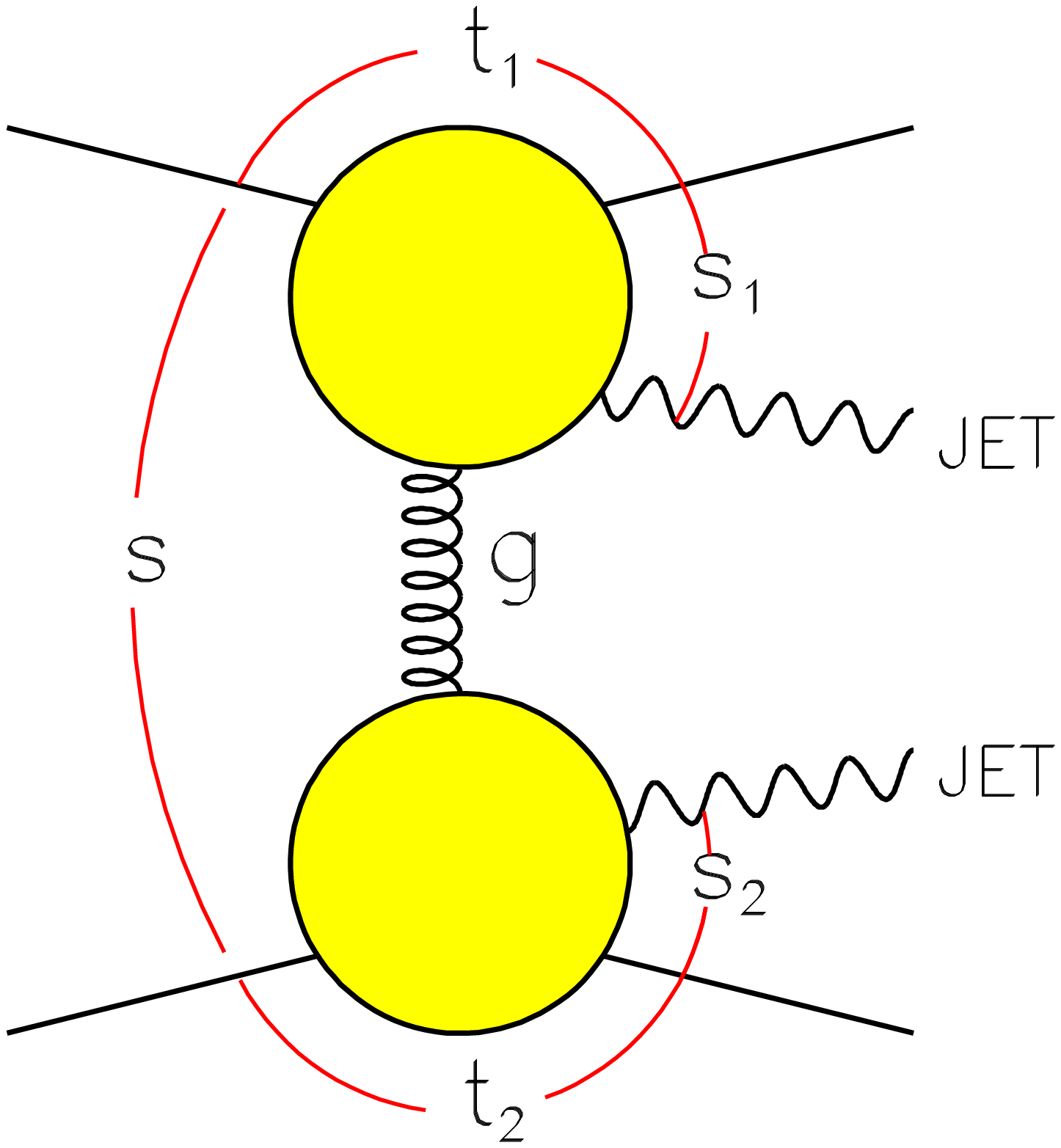}}
 \end{center}
 %\vspace{}
 \caption{Schematic representation of the process $p+p\to p+\mbox{jet}+\mbox{jet}+p$.}
 \end{figure}

  The study of a change of a diffractive pattern
 may be realized as a joint on-line measurement by CMS (jets and
 rapidity gaps) and TOTEM (``diffractive protons'' at Roman Pots)
 \cite{totem}. The dependence
 of a symmetric $(t_1=t_2=t)$ $t$--distribution at two values of $E_T$ is pictured
  at Fig. 7.
\begin{figure}[h]
 \begin{center}
\resizebox{9cm}{!}{\includegraphics{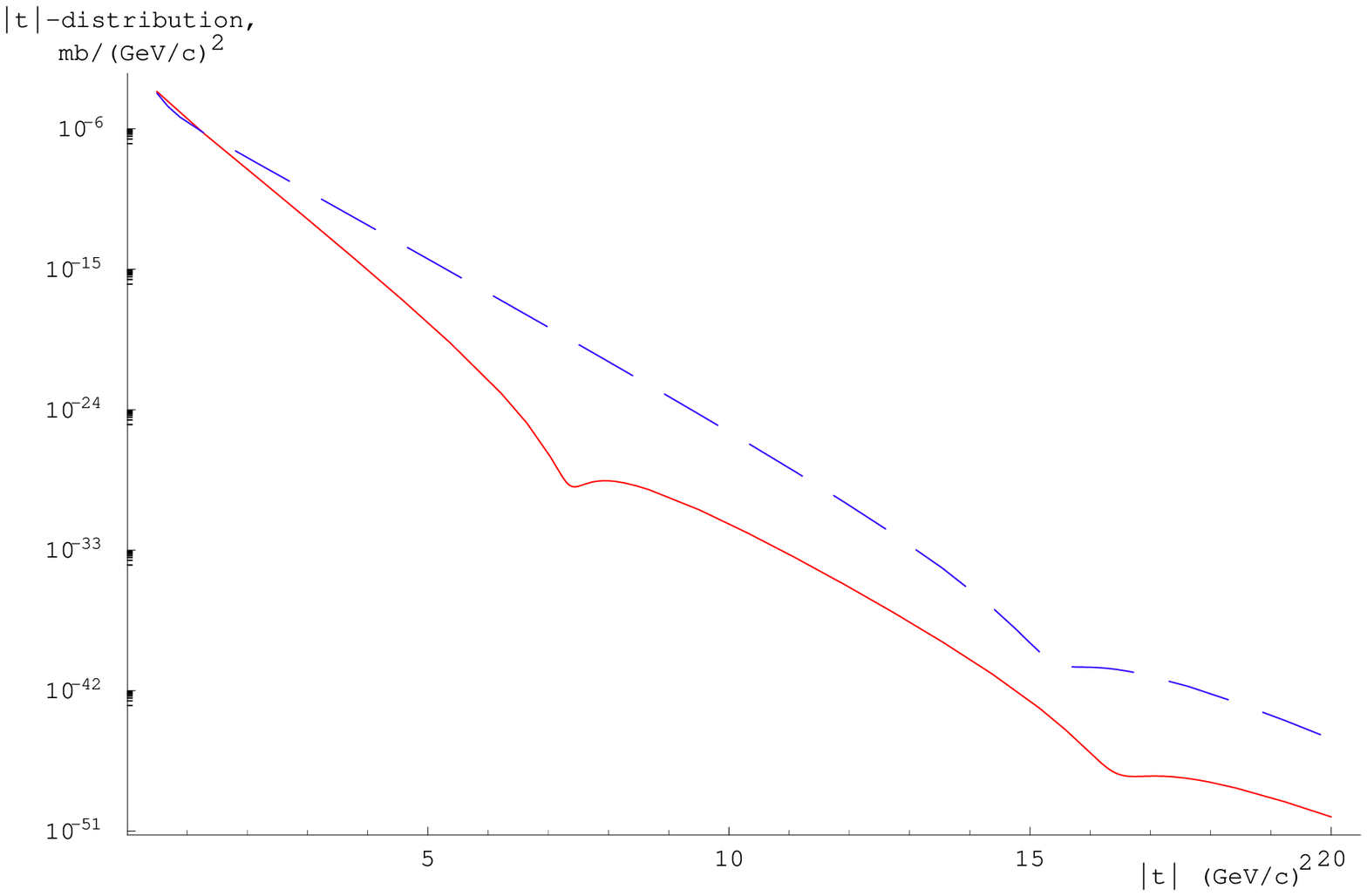}}
\end{center}
 %\vspace{}
 \caption{$s_{1,2}\simeq \sqrt{s}E_t$, $Q^2\simeq 2E_t^2$, solid line
 corresponds to $E_t=10$ GeV, dashed line corresponds to $E_t=100$ GeV;
 $\sqrt{s}=14$ TeV.}
 \end{figure}
The squared sub-energies $s_{1,2}$ are supposed to be in the asymptotical
region.

\section{Conclusion}

The studies of soft interactions at the  LHC energies can lead to
the discoveries of fundamental importance. The evolution
 of hadron scattering with rising energy
 can be described as  transition from the grey to black disc
  and eventually
to black ring with the antishadow scattering mode in the center.
It is worth noting that the \bf{appearance of the antishadow
scattering mode} \rm  at the LHC energy implies a somewhat unusual
scattering picture. At high energies the proton should be
realized as a  loosely bounded composite system and it
appears that this system has a high probability to reinstate
itself only in the central collisions where all of its parts
participate in the coherent interactions. Therefore the central
collisions are responsible for elastic processes while the
peripheral ones where only few parts of weekly bounded protons are
involved result in the production of the secondary particles. This
leads to the peripheral impact parameter profile of the inelastic
overlap function.

 We have to emphasize once again that from the space--time point
 of view high--energy diffractive processes reveal \bf larger and larger
  distances and times \rm which is a real \bf terra incognita \rm
  ``filled'' with still unknown gluon field configurations evidently
   responsible for \bf{confinement dynamics} \rm .

There could be envisaged  various experimental configurations at
the LHC; e.g. soft diffraction  goes well to the interest of the
TOTEM experiment, while hard diffractive final states can be
measured by CMS detector and possible \bf{correlations between the
features of the soft and hard diffractive processes} \rm can be
obtained using  combined measurements of TOTEM and CMS
\cite{petr}.


\newpage
 \small
\begin{thebibliography}{99}
\bibitem{rev}
D. M. Jansen, M. Albrow and R. Brugnera, hep-ph/9905537.
\bibitem{cern}
R. Bonino et al., Phys. Lett. B 211 (1988) 239; A. Brandt et al.,
Nucl. Phys.  B 514 (1998) 3.
\bibitem{h1}
T. Ahmed et al., Phys. Lett. B 348 (1995) 681.
\bibitem{zeus}
M. Derrick et al., Z. Phys. C. 68 (1995) 569.
\bibitem{herdif}
C. Adloff et al., Z. Phys. C. 76 (1997) 613.
\bibitem{tevdif}
L. Alvero et al., Phys. Rev. D 59 (1999) 074022.
\bibitem{phl}
S. M. Troshin and N. E. Tyurin, Phys. Lett. \bf B 316  \rm (1993)
175.
\bibitem{ech}
S. M. Troshin and N. E. Tyurin, Phys. Part. Nucl. 30 (1999) 550.
\bibitem{ltkhs}
A. A. Logunov, V. I. Savrin, N. E. Tyurin and O. A. Khrustalev,
Teor. Mat. Fiz. \bf 6 \rm (1971) 157;
\bibitem{pras}
 P. M. Nadolsky, S. M. Troshin and N. E. Tyurin, Z. Phys.
C \bf  69 \rm (1995)   131 .
\bibitem{laslo}
 P. Desgrolard , L. Jenkovszky, B. Struminsky,  Eur. Phys. J.
  C \bf 11 \rm (1999) 144;\\
P. Desgrolard, hep-ph/0106043.
\bibitem{ezh}
The computer readable files available at http://pdg.lbl.gov.
\bibitem{vels}
J. Velasco , J. Perez-Peraza, A. Gallegos-Cruz, M.
Alvarez-Madrigal, A. Faus-Golfe, A. Sanchez-Hertz,  hep-ph/9910484
\bibitem{bl}
M. M. Block, F. Halzen and T. Stanev, hep-ph/9908222.
\bibitem{will}
C. Adloff et al., Eur. Phys. J. 1998 V. C10\rm , 443 .
\bibitem{angd}
S. M. Troshin and N. E. Tyurin, hep-ph/0008274.
\bibitem{totem}
The TOTEM Collaboration, Technical Proposal CERN/LHCC 99-7, LHCC/P5, 1999.
\bibitem{petr}
V. A. Petrov, talk given at the International Symposium `` LHC
Physics and Detectors'', Dubna, 28-30 June 2000.

\end{thebibliography}
\end{document}